\documentclass[12pt]{article}
\begin{document}

\begin{titlepage}

\centerline{\large \bf Role of relativistic kinematics in describing
two-quark systems}

\vskip 1cm

\centerline{V. Lengyel$^{1}$, V. Rubish$^{1\ast}$, Yu.
Fekete$^{1\dagger}$, A. Shpenik$^{1\ddagger}$, S. Chalupka$^{2}$,
M. Salak$^{3}$}

\vskip .5cm

\centerline{$^{1}$ \sl Uzhgorod State University,} \centerline{\sl
Department of Theoretical Physics,} \centerline{ \sl Voloshin str.
32, 88000 Uzhgorod, Ukraine}

\vskip .3cm

\centerline{$^{2}$ \sl Department of Theoretical Physics and
Geophysics,} \centerline{ \sl P. J. Safarik University, Moyzesova
16,} \centerline{ \sl 041 54 Kosice, Slovak Republic}

\vskip .3cm

\centerline{$^{3}$ \sl Department of Physics, Presov University,}
\centerline{ \sl 17 Novembra, 080 09 Presov, Slovak Republic}

\vskip 1.5cm

\begin{abstract}
An attempt to incorporate relativistic kinematics in the
description of light quark systems is made. It seems that the way
of such incorporation along the suggestion expressed by R. Gaida
and his collaborators is very promising. Comparison of these
results with the experimental data concerning boson mass spectrum
shows that this approach is among the best theoretical
interpretation of the data.
\end{abstract}

\vskip .3cm

\vskip 7cm

\hrule

\vskip .9cm

\noindent
\vfill
$ \begin{array}{ll} ^{\ast}\mbox{{\it e-mail
address:}} &
 \mbox{vrubish@univ.uzhgorod.ua}
\end{array}
$

$ \begin{array}{ll} ^{\dagger}\mbox{{\it e-mail address:}} &
   \mbox{kishs@univ.uzhgorod.ua}
\end{array}
$

$ \begin{array}{ll} ^{\ddagger}\mbox{{\it e-mail address:}} &
   \mbox{shpenik@org.iep.uzhgorod.ua}
\end{array}
$

\vfill
\end{titlepage}
\eject
\baselineskip=14pt

Surprising success of non-relativistic quark models with funnel or
oscillator potential in describing hadron mass spectra for heavy
quark systems had waken the hopes that at least some aspects of
the strong interactions are understood after all. Good insight of
previous works concerning this subject is given in \cite{1,2,3}.
But evidently the picture is not complete until it incorporates
the light quarks $u$, $d$ and $s$. But this means that the problem
is shifting into quantum chromodynamical sector. In this respect
the search of alternative possibilities of incorporating
relativistic kinematics is quite actual. A very promising approach
to the solution of relativistic problem of interaction of two
particles was suggested recently by R. Gaida and his collaborators
\cite{4}-\cite{7}. Their results can be directly applied to our
problem of calculating the mass spectrum of quarkonium as a system
of two-quarks. Somewhat different approach was simultaneously
developed by I. Todorov and P. Bogolyubov \cite{8,9} and later by
E. Predazzi et. al. \cite{10}. In this work we shall investigate
the results concerning the application of their findings to
light-quark systems.

Let us follow first the way, suggested by Predazzi et al
\cite{10}-\cite{12}. Following the ideas expressed in \cite{10,11}
and later developed in \cite{12}, let us start with the classical
expression for the relativistic total energy of two particle
system with masses $m$ and $M$, respectively

\begin{equation}
E=\sqrt{\mathbf{p}_1^2+m^2}+\sqrt{\mathbf{p}_2^2+M^2}  \label{f1}
\end{equation}
were $\mathbf{p}_1=-\mathbf{p}_2\equiv \mathbf{p}$ in the c.m.
system. The main problem here is the mode of inclusion the
interaction between two particles. For equal masses Lucha and
Schoberl \cite{13} write basing on (\ref{f1}) and passing to
quantum mechanics

\begin{equation}
\left( 2\sqrt{\mathbf{p}^2+m^2}+V\right) \Psi \left( \mathbf{r}\right)
=E\Psi \left( \mathbf{r}\right)  \label{f2}
\end{equation}
which they call the spinless Bethe-Salpeter equation. They solve very
cunningly this equation in the configurational momentum representation but
obtain not very good results, which we shall discuss later. But in any case
we can conclude that exactly the way of introducing the interaction is
responsible for their failure. Therefore we propose here another way of
introducing the interaction. Namely, following Predazzi et al \cite{12} we
linearize the expression

\begin{equation}
\left(\frac{E^2+m^2-M^2}{2E}\right)^2=\mathbf{p}^2+m^2 \label{f3}
\end{equation}
which follows from (\ref{f1}) after simple algebraic transformations, and
obtain

\begin{equation}
\left( \frac{E^2+m^2-M^2}{2E}\right) \Psi \left( \mathbf{r}\right)
=(\mathbf{\alpha }c\mathbf{p+\beta }mc^2+\widetilde{V})\Psi \left(
\mathbf{r}\right)  \label{f4}
\end{equation}
where $\mathbf{\alpha }$ and $\mathbf{\beta }$ are the usual Dirac matrices,
$\Psi$ is the four- component wave function for which we shall use
two-component representation
\begin{equation}
\Psi \left( \mathbf{r}\right) =\left(
\begin{array}{c}
\varphi \left( \mathbf{r}\right) \\
\chi \left( \mathbf{r}\right)
\end{array}
\right) .  \label{f5}
\end{equation}
In the general case we would have on this stage to decide what kind of
Lorentz-transform properties we shall ascribe to the interaction.

In general the interaction can transform either like Lorentz-scalar (like
the mass) or to be a 4-th component of Lorentz-vector, i.e. transform like
energy.Then we shall consider the interaction $\widetilde{V}$ as a mixture

\begin{equation}
\widetilde{V}=\mathbf{\beta }S\cdot \left( 1-\varepsilon \right) +
\mathbf{I}V\cdot \varepsilon  \label{f6}
\end{equation}
where $\varepsilon$ is some mixing parameter. In what follows we shall
simplify the expression for $\widetilde{V}$ by taking $\varepsilon=1/2$
(which is suggested by experimental data (see,e. g. \cite{14} and also \cite
{15}) what means that (\ref{f6}) can be written in the form

\begin{equation}
\widetilde{V}=\frac 12(\mathbf{\beta +I)}\left( V+S\right) .
\label{f7}
\end{equation}

This form is chosen in order to obtain simple non-relativistic
result $\widetilde{V}=V$ if there is no difference between $S$ and
$V$. Such potential was introduced previously by V. Kukulin, M.
Moshinsky et. al. \cite{16,17} and was called an averaged
potential, which allows to reduce the system of Dirac equations to
a single relativistic oscillator equation.In this, particular case
when the mixture of scalar $S$ and vector $V$ part of potentials
are equal we shall consider the following possibilities

\begin{equation}
S\left( r\right) +V\left( r\right) \equiv
\widetilde{V}=-\frac{\alpha _{s}}{r}+A\cdot r^{2}+V_{0},
\label{f8}
\end{equation}

\begin{equation}
S\left( r\right) +V\left( r\right) \equiv
\widetilde{V}=-\frac{\alpha _s}r+k\cdot r+V_{0},  \label{f9}
\end{equation}

\begin{equation}
S\left( r\right) +V\left( r\right) \equiv
\widetilde{V}=\frac{g^2}{6\pi \mu } \left( 1-e^{-\mu r}\right)
-\frac{16\pi }{25}\frac{e^{-kr}}{r\cdot \ln \left( b+\left( \frac
1{\Lambda r}\right) ^2\right) }+V_{0}.  \label{f10}
\end{equation}

Today all of these potentials are used to describe the
quark-antiquark interaction. The discussion concerning the
advantages and handicapes of these potentials is given in
\cite{18}. Our aim is to apply these potentials to describing the
meson mass spectra with relativistic kinematics which is built
into equation (\ref{f4}). In comparison to \cite{18} the search of
the best description by minimizing $\chi^2$ will be given. For an
averaged potential (\ref{f7}) the equation (\ref{f4}) reduces to
single equation for the large wave function $\varphi (\mathbf{r})$

\begin{equation}
\left( E-V\right) ^2\varphi \left( \mathbf{r}\right) =\left[
4\mathbf{p}^2+4m^2+4mS+S^2\right] \varphi \left( \mathbf{r}\right)
=\left[ 4\mathbf{p}^2+4\left( m+\frac S2\right) ^2\right] \varphi
\left( \mathbf{r}\right).  \label{f11}
\end{equation}

For this propose we have to solve numerically the following equation

\begin{equation}
\lbrack \mathbf{p}^2+\left( \frac E2+m\right) \cdot
\frac{\widetilde{V}} 2-\left( \frac{E^2}4-m^2\right) ]\varphi
\left( \mathbf{r}\right)=0. \label{f12}
\end{equation}

Transferring to operators and carrying out the substitutions for unknown
function

\begin{equation}
\varphi \left( \mathbf{r}\right) =\frac{\Phi \left(
\mathbf{r}\right) }{\mathbf{r}}  \label{f13}
\end{equation}
one obtains the equation

\begin{equation}
\left[ \frac{d^2}{dr^2}-\frac{l(l+1)}{r^2}-\left( \frac E2+m\right) \cdot
\frac{\widetilde{V}}2+\frac{E^2}4-m^2\right] \Phi \left( \mathbf{r}\right)
=0.  \label{f14}
\end{equation}

In a simple approximation

\begin{equation}
S+V=\frac 12\left( Ar^2+V_0\right)  \label{f15}
\end{equation}
one obtains

\begin{equation}
\lbrack \mathbf{p}^2+\left( \frac E2+m\right) \left( \frac 12Ar^2+\frac
12V_0\right) -\left( \frac{E^2}4-m^2\right) ]\varphi \left( \mathbf{r}
\right) =0  \label{f16}
\end{equation}
which leads to the equation for relativistic isotropic oscillator

\begin{equation}
\left[ \frac{d^2}{dr^2}-\frac{l(l+1)}{r^2}-\left( \frac E2+m\right) \cdot
\frac 12Ar^2+\frac{E^2}4-m^2-\left( \frac E2+m\right) \cdot \frac
12V_0\right] \Phi \left( \mathbf{r}\right) =0.  \label{f17}
\end{equation}

Now with the standard change of variables

\begin{equation}
\sqrt{\left( \frac E2+m\right) \frac 12A}\cdot r^2=x^2  \label{f18}
\end{equation}
one obtains

\begin{equation}
\left[ \frac{d^2}{dx^2}-\frac{l(l+1)}{x^2}-x^2+\frac{E^2/4-m^2-\frac
12V_0\left( \frac E2+m\right) }{\sqrt{\left( \frac E2+m\right) \frac A2}}
\right] \Phi \left( \mathbf{r}\right) =0.  \label{f19}
\end{equation}

The physical solution of (\ref{f7}) is satisfied, when

\begin{equation}
\frac{E^2}4-m^2=\sqrt{\frac A2\cdot \frac{E+2m}2}\cdot \left( 4N+2l+3\right)
+\frac{V_0\left( E/2+m\right) }2  \label{f20}
\end{equation}
or

\begin{equation}
\frac E2-m=\sqrt{\frac A{E+2m}}\left( 4N+2l+3\right) +\frac{V_0}2
\label{f21}
\end{equation}

In the non-relativistic limit when $E\sim 2m$ one has

\begin{equation}
E\approx 2m+\sqrt{\frac Am}\left( 4N+2l+3\right) +V_0  \label{f22}
\end{equation}
in full accordance with the non-relativistic case (see,e.g.\cite{18}).

Actually for large total energies $E$ from (\ref{f21}) it follows $E^2\sim
l^{4/3}$ i.e. almost linear Regge-trajectory as it should be from general
considerations. As we shall see the application of other variants of
potentials gives even better results.

As we have mentioned earlier, recently very promising $\pi$
approach to the problem of relativistic description of
many-particle system was elaborated along with the ideas presented
by Gaida in \cite{4} and elaborated by Tretyak, Spytko and
Duviryak \cite{5,6,7}. They used the Weyl quantization method and
succeeded in solving the problem for relativistic oscillator
between two particles.

Considering oscillator-type interaction

\begin{equation}
V=\omega ^2\mathbf{p}_1\mathbf{p}_2r^2  \label{f23}
\end{equation}
where

\begin{equation}
\mathbf{p}_1\mathbf{p}_2=\frac{m_{RED}}2=\frac{m_q}4  \label{f24}
\end{equation}
the non-relativistic approximation they would obtain for mass of
two-particle system

\begin{equation}
M=\left\{ \left[ 2m_q+\frac \omega 2\left( 4N+2l+3\right) \right]
^2+\frac{\omega ^2}4\right\} ^{1/2}+V_0  \label{f25}
\end{equation}
or

\begin{equation}
M=\sqrt{\left( \sum m_q+\sqrt{\frac A{m_q}}\cdot \left( 4N+2l+3\right)
\right) ^2+\frac A{m_q}}+V_0  \label{f26}
\end{equation}
if we express the (\ref{f26}) in form of string tension $A$ and generalize
their results to our boundary condition of isotropic oscillator.

Exactly such approximation was taken in \cite{18}. But if we suggest that in
general

\begin{equation}
\mathbf{p}_1\mathbf{p}_2=\frac{E+m_q}8  \label{f27}
\end{equation}
then Lvov group results

\begin{equation}
M=\left\{ \left[ 2m_q+\sqrt{\frac{2A}{E+m_q}}\cdot \left( 4N+2l+3\right)
\right] ^2+\frac{2A}{E+m_q}\right\} ^{1/2}+V_0  \label{f28}
\end{equation}
will resemble very much our relativistic approach. The results of
calculations according to (\ref{f28}) are given below. The
parameters are taken to be $A= 0.01$ $GeV^{3}$, $V_0=-0.436$ $GeV$
certainly this approximation is valid if one considers the energy
dependence of $\mathbf{p}_{i}$ only on the last stage of
calculations.

It is interesting that similar to (\ref{f28}) result for $\mathbf{{M}^2}$-
operator was obtained in relativistic approach by Ishida-Oda based on
special assumption of covariant relativistic approach. Even the numerical
values of parameters of Ishida-Oda \cite{19} are close to the result of (\ref
{f28}). Their value $A=0.05 GeV^3$ which is of the same order as our $A$.
More precise comparison is impossible because of ambiguity of other
parameters.

Usually such potential models like we have used here are called na\"\i ve
quark model. But exactly our model is not-so na\"\i ve. Firstly, the
relativistic kinematics not only renders it more complicated, but shows the
possible way of building the model of interaction of two relativistic
particles. Secondly the potentials like (\ref{f8}-\ref{f10}) incorporate
asymptotic freedom the strong coupling constant $\alpha _s$ was calculated
according to the classical expression

\begin{equation}
\alpha _s\left( r\right) =\frac{12\pi }{33-2N}\cdot \frac 1{\ln \left( \frac
1{r^2\widetilde{\Lambda }^2}\right) }  \label{f29}
\end{equation}
where $\widetilde{\Lambda }$ was taken to be equal to
$\widetilde{\Lambda }=0.14GeV$. And lastly the model allows to
include spin-spin interaction either by passing to Breit-Fermi
equation or by using Dirac equation straightly. In the Table 1
$\alpha _s$ is taken exactly according to (\ref {f29}).

It is interesting to note that the definition of mass can be given in a
different way. Considering one particle as moving in the field of another
and vice versa and adding the obtained masses we obtain the results with
parameters which correspond very closely to our Table 1. Namely one can
write down the Dirac equation for one particle moving in the outer field,
reduce it to the equation for large component $\chi \left( \mathbf{r}\right)
$ and obtain \cite{20}

\begin{equation}
\left( E^2-m^2\right) \chi \left( \mathbf{r}\right) =\mathbf{p}^2\chi \left(
\mathbf{r}\right) +\left( E+m\right) V\chi \left( \mathbf{r}\right) .
\label{f30}
\end{equation}

Applying the virial theorem to this equation

\begin{equation}
\left\langle \chi (y)\left| \mathbf{p}^2(y)\right| \chi (y)\right\rangle
=\frac 12\left\langle \chi (y)\left| y\frac{\partial V}{\partial y}\right|
\chi (y)\right\rangle =\left\langle \chi (y)\left| y^2\right| \chi
(y)\right\rangle  \label{f31}
\end{equation}

Combining (30) and (31) for oscillator interaction one obtains

\begin{equation}
E^2-m^2=2\sqrt{C\left( E+m\right) }B_{Nl}+\left( E+m\right) V_{0}
\label{f32}
\end{equation}
where

\begin{equation}
B_{Nl}=\left\langle \chi (y)\left| y^2\right| \chi (y)\right\rangle =2N+l+3/2
\label{f33}
\end{equation}

This expression on resembles very much the (\ref{f28}). So we are left with
the five most realistic from our point view possibilities, namely, of
calculating masses according to (\ref{f8}), (\ref{f9}), (\ref{f10}), (\ref
{f26}), (\ref{f28}).

To obtain the masses of multi-quark system according to these expressions,
it is necessary to define the values of parameters. The masses of quarks
were taken to be as usually in quark models $m_u=0.33 GeV$, $m_c =
1.675-1.75 GeV$, $m_b = 5.05-5.1 GeV$. Parameters for different potentials
are shown in Table 1.

\medskip\

Table 1. The parameter values for different potentials

\begin{tabular}{|cccccc|}
\hline \multicolumn{1}{|c|}{$Potential$} &
\multicolumn{1}{c|}{$\alpha _u$} & \multicolumn{1}{c|}{$\alpha
_c$} & \multicolumn{1}{c|}{$\alpha _b$} & \multicolumn{1}{c|}{$k,$
$A$} & $V_0,$ $GeV$ \\ \hline \multicolumn{1}{|c|}{$(\ref{f8})$} &
\multicolumn{1}{c|}{$0.5$} & \multicolumn{1}{c|}{$0.325$} &
\multicolumn{1}{c|}{$0.3$} & \multicolumn{1}{c|}{
\begin{tabular}{c}
$k=0.27$ \\ $(GeV^{2})$
\end{tabular}
} & $-0.8356$ \\ \hline \multicolumn{1}{|c|}{$(\ref{f9})$} &
\multicolumn{1}{c|}{$0.5$} & \multicolumn{1}{c|}{$0.386$} &
\multicolumn{1}{c|}{$0.3$} & \multicolumn{1}{c|}{
\begin{tabular}{c}
$A=0.01$ \\ $(GeV^3)$
\end{tabular}
} &
\begin{tabular}{c}
$(-0.436)...$ \\ $...(-0.527)$
\end{tabular}
\\ \hline
\multicolumn{1}{|c|}{$(\ref{f10})$} & \multicolumn{1}{c|}{
\begin{tabular}{c}
$\frac{g^2}{6\pi }=0.3795$ \\ $(GeV^{2})$
\end{tabular}
} & \multicolumn{1}{c|}{
\begin{tabular}{c}
$\mu =0.054$ \\ $(GeV)$
\end{tabular}
} & \multicolumn{1}{c|}{
\begin{tabular}{c}
$K=0.75$ \\ $(GeV)$
\end{tabular}
} & \multicolumn{1}{c|}{
\begin{tabular}{c}
$\Lambda =0.35$ \\ $(GeV)$ \\ $b=4$
\end{tabular}
} & $-1.103$ \\ \hline
\end{tabular}

\newpage\

Table 2.Variation of $\chi ^2$ for different parameters of potential (\ref
{f9}).

\begin{tabular}{|ccccc|}
\hline
\multicolumn{1}{|c|}{
\begin{tabular}{c}
$k$ \\ $(GeV^{2})$
\end{tabular}
} & \multicolumn{1}{c|}{$\chi _{u\overline{u}}^2$} &
\multicolumn{1}{c|}{$\chi _{c\overline{c}}^2$} &
\multicolumn{1}{c|}{$\chi _{b\overline{b}}^2$} &
\begin{tabular}{c}
$V_0$ \\ $(GeV)$
\end{tabular}
\\ \hline
\multicolumn{1}{|c|}{$0.1826$} & \multicolumn{1}{c|}{$514$} &
\multicolumn{1}{c|}{$1.15\cdot 10^5$} & \multicolumn{1}{c|}{$1.3\cdot 10^4$}
& $-0.628$ \\ \hline
\multicolumn{1}{|c|}{$0.25$} & \multicolumn{1}{c|}{$98$} &
\multicolumn{1}{c|}{$6.2\cdot 10^3$} & \multicolumn{1}{c|}{$2\cdot 10^4$} & $%
-0.791$ \\ \hline
\multicolumn{1}{|c|}{$0.26$} & \multicolumn{1}{c|}{$67$} &
\multicolumn{1}{c|}{$2.2\cdot 10^4$} & \multicolumn{1}{c|}{$1.6\cdot 10^4$}
& $-0.813$ \\ \hline
\multicolumn{1}{|c|}{$0.27$} & \multicolumn{1}{c|}{$43.6$} &
\multicolumn{1}{c|}{$3\cdot 10^4$} & \multicolumn{1}{c|}{$2.9\cdot 10^3$} & $%
-0.835$ \\ \hline
\multicolumn{1}{|c|}{$0.29$} & \multicolumn{1}{c|}{$12.6$} &
\multicolumn{1}{c|}{$7\cdot 10^4$} & \multicolumn{1}{c|}{$8.5\cdot 10^3$} & $%
-0.878$ \\ \hline
\multicolumn{1}{|c|}{$0.305$} & \multicolumn{1}{c|}{$1.8$} &
\multicolumn{1}{c|}{$1.36\cdot 10^5$} & \multicolumn{1}{c|}{$4.3\cdot 10^4$}
& $-0.908$ \\ \hline
\end{tabular}

\medskip\

The value of $V_{0}$ reflects the fact that potential is the
Fourier-transform of scattering amplitude $V_{0}$ being the
constant of interaction. Gromes $\left[ 21\right]$ has evaluated
this constant for linear confinement and obtained the value of
$V_0$

\begin{equation}
V_0\simeq -2\sqrt{k}\cdot e^{-\left( \gamma -0.5\right) }  \label{f34}
\end{equation}
where $k$ is the string tension $\gamma =0.57721...$ is Euler-MacLoraint
constant. According to our values of $k=0.18-0.305GeV^2$. According to (\ref
{f34}) has to vary within the limits $V_0=-\left( 0.77\div 1.02\right) GeV$
which is quite close to values, cited in the table (\ref{f2}). Let us stress
that the set values are not colour-dependent, which, reduces the number of
adjustable parameters.

The result of calculation together with experimental data are
shown in Tables (\ref{f3}-\ref{f6}). Experimental values were
taken from \cite{22}. For choosing the parameters the minimum of
$\chi ^2-$criterion was used, with the definition of $\chi ^2$
given in \cite{18}. In this definition $\mathbf{N}$ is the number
of meson masses, $\mathbf{n-}$ is the number of parameters (in our
case we considered them to be equal to two namely confinement
parameter and $V_0$), $\Delta $ is the experimental error in
definition of experimental mass $M_{EXP}$ of two-quark system
\cite{22} since $V_0$ was chosen to match the experimental value
of ground-state mass we are actually left only which one
adjustable parameter $A$ (or $k$). Since we do not include
LS-forces we had to take the average center of gravity (COG) value
of $P$-resonances, which was calculated according to formula

\begin{equation}
M_{COG}=\frac{\sum_{j}\left( 2j+1\right) \cdot
M_{j}}{\sum_{j}\left( 2j+1\right)}.  \label{f35}
\end{equation}

As one can see from both radial and orbital excitation calculation the
variant which incorporates in one or another way the relativistic kinematics
give better description of Regge-trajectories, which, are believed to be
linear in $l$ for $M^2.$

It is well known that even in non-relativistic limit one can obtain good
description of $u\overline{u}$-$d\overline{d}$ -systems at the account of
spoiling the $c\overline{c}$ of $b\overline{b}$ -description.

Indeed in the Table 2 we show $\chi ^2$ obtained for Eichten
parameters. Due to a high precision of defining $J/\Psi $ and
$\Upsilon$-mesons the large $\chi ^2$ were considered good for
these mesons and bad for $\rho $-meson trajectory. On other hand
the non-relativistic Badalyan \cite{23} results are good for
$u\overline{u}$, but bad for $J/\Psi$. Fabre \cite{24} in order to
obtain good results for light quarkonium had to change the very
potential. Instead of this situation our way or incorporating the
relativistic kinematics we obtain good results for all data. The
table 2 demonstrates this statement. Table 3 contains the
comparison of different potentials. The discussion concerning the
choice of potentials is given in \cite{18}. We shall choose in
what follows the Cornell-potential (\ref{f9}) which seems to be
preferable though potential (\ref{f10}) is also quite good. The
parameters $\alpha _s$ here are taken from table 1. The
$u\overline{u}$-data are fantastically good, but $c\overline{c}$
and $b\overline{b}$-data could be better.Therefore in tables 4-6
we give the results for $k=0.29 GeV^2$. We consider these results
as the best. It is interesting that the values which give these
best results are close to those of Lucha and Sch\"oberl
\cite{3}.We want to stress that all the above results are obtained
by numerical solution of (\ref{f14}). The last columns in
$\left(4-6\right)$ are calculated according to (\ref{f28} ). With
the choice $A=0.071 GeV^{3}$, $V_0=-1.0077 GeV$, $m_u=0.33 GeV$,
$m_c=1.75 GeV$, $m_b=5.13 GeV$ the results are quite comparable
with the other entries. But still we have to conclude that pure
oscillator potential is too rough to give final result. The use of
more sophisticated potential is to be taken hear too. But it
demonstrates nicely that inclusion of relativistic kinematics is
very crucial.

We would like to call the attention to one interesting feature of
relativistic models, namely that the slope of linear (or close to linear)
Regge-trajectory in this case is constant, while in nonlinear models it is
neither constant nor linear. Roughly experiments show this slope to be equal
$\sim 1.2 GeV^{2}$. In our cases it varies from $1.15$ to $2.5 GeV^{2}$ for
different cases. As Tutik et \cite{25} have indicated the Regge trajectories
for low-lying states coincide while for large values of orbital momentum $l$
the screened potential (\ref{f10}) leads a limited Regge-trajectory in
contrast to infinitely rising trajectories for other potentials like (\ref
{f8}) or (\ref{f9}).

\medskip\

Table 3. Mass spectrum of $u\overline{u}$-system with some realistic
potentials.

\begin{tabular}{|cccccc|}
\hline \multicolumn{1}{|c|}{$State$} &
\multicolumn{1}{c|}{$designation$} & \multicolumn{1}{c|}{
\begin{tabular}{c}
$M_{EXP}$ \\ $GeV$
\end{tabular}
} & \multicolumn{1}{c|}{
\begin{tabular}{c}
$M_{TH} (\ref{f8})$ \\ $GeV$
\end{tabular}
} & \multicolumn{1}{c|}{
\begin{tabular}{c}
$M_{TH} (\ref{f9})$ \\ $GeV$
\end{tabular}
} &
\begin{tabular}{c}
$M_{TH} (\ref{f10})$ \\ $GeV$
\end{tabular}
\\ \hline
\multicolumn{1}{|c|}{$1S$} & \multicolumn{1}{c|}{$\rho$ $
1^{+}(1^{--})$} & \multicolumn{1}{c|}{$0.768\pm 0.0005$} &
\multicolumn{1}{c|}{$0.768$} & \multicolumn{1}{c|}{$0.768$} &
$0.768$ \\ \hline \multicolumn{1}{|c|}{$1P$} &
\multicolumn{1}{c|}{$^3P_{COG}^{*}$} &
\multicolumn{1}{c|}{$1.262\pm 0.03$} & \multicolumn{1}{c|}{$1.26$}
& \multicolumn{1}{c|}{$1.296$} & $1.3$ \\ \hline
\multicolumn{1}{|c|}{$2S$} & \multicolumn{1}{c|}{$\rho$ $
1^{+}(1^{--})$} & \multicolumn{1}{c|}{$1.465\pm 0.025$} &
\multicolumn{1}{c|}{$1.609$} & \multicolumn{1}{c|}{$1.573$} &
$1.561$ \\ \hline \multicolumn{1}{|c|}{$2P$} &
\multicolumn{1}{c|}{$a_{2}$ $1^{-}(2^{++})$} &
\multicolumn{1}{c|}{$1.935\pm 0.015$} &
\multicolumn{1}{c|}{$2.003$} & \multicolumn{1}{c|}{$1.932$} &
$1.889$ \\ \hline \multicolumn{1}{|c|}{$3S$} &
\multicolumn{1}{c|}{$\rho$ $1^{+}(1^{--})$} &
\multicolumn{1}{c|}{$2.15\pm 0.01$} & \multicolumn{1}{c|}{$2.3$} &
\multicolumn{1}{c|}{$2.155$} & $2.103$ \\ \hline
\multicolumn{1}{|c|}{$1D$} & \multicolumn{1}{c|}{$\rho_{3}$ $
1^{+}(3^{--})$} & \multicolumn{1}{c|}{$1.691\pm 0.013$} &
\multicolumn{1}{c|}{$1.665$} & \multicolumn{1}{c|}{$1.689$} &
$1.687$ \\ \hline \multicolumn{1}{|c|}{$2D$} &
\multicolumn{1}{c|}{$\rho_{3}$ $1^{+}(3^{--})$} &
\multicolumn{1}{c|}{$2.25\pm 0.01$} & \multicolumn{1}{c|}{$2.348$}
& \multicolumn{1}{c|}{$2.235$} & $2.1845$ \\ \hline
\multicolumn{1}{|c|}{$1F$} & \multicolumn{1}{c|}{$a_{4}$ $
1^{-}(4^{++})$} & \multicolumn{1}{c|}{$2.037\pm 0.036$} &
\multicolumn{1}{c|}{$2.03$} & \multicolumn{1}{c|}{$2.021$} &
$2.003$ \\ \hline \multicolumn{1}{|c|}{$1G$} &
\multicolumn{1}{c|}{$\rho_{5}$ $1^{+}(5^{--})$} &
\multicolumn{1}{c|}{$2.350\pm 0.015$} &
\multicolumn{1}{c|}{$2.366$} & \multicolumn{1}{c|}{$2.312$} &
$2.275$ \\ \hline \multicolumn{1}{|c|}{$1H$} &
\multicolumn{1}{c|}{$a_{6}$ $1^{-}(6^{++})$} &
\multicolumn{1}{c|}{$2.45\pm 0.13$} & \multicolumn{1}{c|}{$2.685$}
& \multicolumn{1}{c|}{$2.576$} & $2.514$ \\ \hline
\multicolumn{1}{|c|}{$\chi ^2$} & \multicolumn{1}{c|}{$-$} &
\multicolumn{1}{c|}{$-$} & \multicolumn{1}{c|}{$53.3$} &
\multicolumn{1}{c|}{$1.8$} & $19.6$ \\ \hline
\end{tabular}

\newpage\

Table 4. The best mass spectrum of $u\overline{u}$-system with
Cornell-potential (\ref{f9})

in formula (\ref{f28}).

\begin{tabular}{|ccccc|}
\hline \multicolumn{1}{|c|}{$State$} &
\multicolumn{1}{c|}{$designation$} & \multicolumn{1}{c|}{
\begin{tabular}{c}
$M_{EXP}$ \\ $GeV$
\end{tabular}
} & \multicolumn{1}{c|}{
\begin{tabular}{c}
$M_{TH} (\ref{f9})$ \\ $GeV$
\end{tabular}
} &
\begin{tabular}{c}
$M_{TH} (\ref{f28})$ \\ $GeV$
\end{tabular}
\\ \hline
\multicolumn{1}{|c|}{$1S$} & \multicolumn{1}{c|}{$\rho$
$1^{+}(1^{--})$} & \multicolumn{1}{c|}{$0.768\pm 0.0005$} &
\multicolumn{1}{c|}{$0.768$} & $0.768$ \\ \hline
\multicolumn{1}{|c|}{$1P$} & \multicolumn{1}{c|}{$^3P_{COG}^{*}$}
& \multicolumn{1}{c|}{$1.262\pm 0.03$} &
\multicolumn{1}{c|}{$1.281$} & $1.198$
\\ \hline
\multicolumn{1}{|c|}{$2S$} & \multicolumn{1}{c|}{$\rho$
$1^{+}(1^{--})$} & \multicolumn{1}{c|}{$1.465\pm 0.025$} &
\multicolumn{1}{c|}{$1.551$} & $1.577$ \\ \hline
\multicolumn{1}{|c|}{$2P$} & \multicolumn{1}{c|}{$a_{2}
1^{-}(2^{++})$} & \multicolumn{1}{c|}{$1.935\pm 0.015$} &
\multicolumn{1}{c|}{$1.902$} & $1.923$ \\ \hline
\multicolumn{1}{|c|}{$3S$} & \multicolumn{1}{c|}{$\rho$
$1^{+}(1^{--})$} & \multicolumn{1}{c|}{$2.15\pm 0.01$} &
\multicolumn{1}{c|}{$2.12$} & $ 2.244$ \\ \hline
\multicolumn{1}{|c|}{$1D$} & \multicolumn{1}{c|}{$\rho _{3}$
$1^{+}(3^{--})$} & \multicolumn{1}{c|}{$1.691\pm 0.013$} &
\multicolumn{1}{c|}{$1.665$} & $1.577$ \\ \hline
\multicolumn{1}{|c|}{$2D$} & \multicolumn{1}{c|}{$\rho _{3}$
$1^{+}(3^{--})$} & \multicolumn{1}{c|}{$2.25\pm 0.01$} &
\multicolumn{1}{c|}{$ 2.197$} & $2.244$ \\ \hline
\multicolumn{1}{|c|}{$1F$} & \multicolumn{1}{c|}{$a_{4}$
$1^{-}(4^{++})$} & \multicolumn{1}{c|}{$2.037\pm 0.036$} &
\multicolumn{1}{c|}{$1.987$} & $1.923$ \\ \hline
\multicolumn{1}{|c|}{$1G$} & \multicolumn{1}{c|}{$\rho _{5}$
$1^{+}(5^{--})$} & \multicolumn{1}{c|}{$2.350\pm 0.015$} &
\multicolumn{1}{c|}{$2.278$} & $2.244$ \\ \hline
\multicolumn{1}{|c|}{$1H$} & \multicolumn{1}{c|}{$a_{6}$
$1^{-}(6^{++})$} & \multicolumn{1}{c|}{$2.45\pm 0.13$} &
\multicolumn{1}{c|}{$2.529$} & $2.547 $ \\ \hline
\multicolumn{1}{|c|}{$\chi ^2$} & \multicolumn{1}{c|}{$-$} &
\multicolumn{1}{c|}{$-$} & \multicolumn{1}{c|}{$12.6$} & $35.68$
\\ \hline
\end{tabular}

\medskip\

Table 5. The same for $c\overline{c}$-system.

\begin{tabular}{|ccccc|}
\hline \multicolumn{1}{|c|}{$State$} &
\multicolumn{1}{c|}{$designation$} & \multicolumn{1}{c|}{
\begin{tabular}{c}
$M_{EXP}$ \\ $GeV$
\end{tabular}
} & \multicolumn{1}{c|}{
\begin{tabular}{c}
$M_{TH} (\ref{f9})$ \\ $GeV$
\end{tabular}
} &
\begin{tabular}{c}
$M_{TH} (\ref{f28})$ \\ $GeV$
\end{tabular}
\\ \hline
\multicolumn{1}{|c|}{$1S$} & \multicolumn{1}{c|}{$J/\Psi$
$0^{-}(1^{--})$} & \multicolumn{1}{c|}{$3.096\pm 0.00009$} &
\multicolumn{1}{c|}{$3.07$} & $3.014$ \\ \hline
\multicolumn{1}{|c|}{$1P$} & \multicolumn{1}{c|}{$\chi _{cl}$
$0^{+}(1^{++})$} & \multicolumn{1}{c|}{$3.51\pm 0.00012$} &
\multicolumn{1}{c|}{$3.5118$} & $3.331$ \\ \hline
\multicolumn{1}{|c|}{$1D$} & \multicolumn{1}{c|}{$\Psi$
$?^{?}(1^{--})$ } & \multicolumn{1}{c|}{$3.770\pm 0.0025$} &
\multicolumn{1}{c|}{$3.837$} & $3.632$ \\ \hline
\multicolumn{1}{|c|}{$2S$} & \multicolumn{1}{c|}{$\Psi$
$0^{-}(1^{--})$} & \multicolumn{1}{c|}{$3.688\pm 0.0001$} &
\multicolumn{1}{c|}{$3.732$} & $3.632$ \\ \hline
\multicolumn{1}{|c|}{$2D$} & \multicolumn{1}{c|}{$\Psi$
$?^{?}(1^{--})$} & \multicolumn{1}{c|}{$4.159\pm 0.02$} &
\multicolumn{1}{c|}{$4.296$} & $4.195$ \\ \hline
\multicolumn{1}{|c|}{$3S$} & \multicolumn{1}{c|}{$\Psi$
$?^{?}(1^{--})$} & \multicolumn{1}{c|}{$4.04\pm 0.01$} &
\multicolumn{1}{c|}{$4.227$} & $4.195$ \\ \hline
\multicolumn{1}{|c|}{$3D$} & \multicolumn{1}{c|}{$\Psi$
$?^{?}(1^{--})$} & \multicolumn{1}{c|}{$4.415\pm 0.006$} &
\multicolumn{1}{c|}{$4.692$} & $4.717$ \\ \hline
\multicolumn{1}{|c|}{$\chi ^2$} & \multicolumn{1}{c|}{$-$} &
\multicolumn{1}{c|}{$-$} & \multicolumn{1}{c|}{$7\cdot 10^4$} &
$8.4\cdot 10^5$ \\ \hline
\end{tabular}

\medskip\

Table 6. The same for $b\overline{b}$-system.

\begin{tabular}{|c|c|c|c|c|}
\hline $State$ & $designation$ &
\begin{tabular}{c}
$M_{EXP}$ \\ $GeV$
\end{tabular}
&
\begin{tabular}{c}
$M_{TH} (\ref{f9})$ \\ $GeV$
\end{tabular}
&
\begin{tabular}{c}
$M_{TH} (\ref{f28})$ \\ $GeV$
\end{tabular}
\\ \hline
$1S$ & $\Upsilon$ $?^{?}\left( 1^{--}\right) $ & $9.460\pm
0.00022$ & $9.479$ & $9.542$ \\ \hline $1P$ & $^3P_{COG}^{*}$ &
$9.892\pm 0.0007$ & $9.883$ & $9.741$ \\ \hline $2S$ & $\Upsilon$
$?^{?}(1^{--})$ & $10.023\pm 0.00031$ & $10.037$ & $9.932$ \\
\hline $2P$ & $^3P_{COG}^{*}$ & $10.268\pm 0.00057$ & $10.299$ &
$10.121$ \\ \hline $3S$ & $\Upsilon$ $?^{?}(1^{--})$ & $10.355\pm
0.0005$ & $10.433$ & $10.308$ \\ \hline $4S$ & $\Upsilon$
$?^{?}(1^{--})$ & $10.58\pm 0.0035$ & $10.776$ & $10.675$ \\
\hline $5S$ & $\Upsilon$ $?^{?}(1^{--})$ & $10.865\pm 0.008$ &
$11.062$ & $11.034$ \\ \hline $6S$ & $\Upsilon$ $?^{?}(1^{--})$ &
$11.019\pm 0.008$ & $11.332$ & $11.385$ \\ \hline $\chi ^2$ & $-$
& $-$ & $8.5\cdot 10^3$ & $7.4\cdot 10^4$ \\ \hline
\end{tabular}

\newpage\

\end{document}